\documentclass{ws-procs9x6}
\usepackage{subfigure}
\usepackage{yfonts}
\usepackage{longtable}

\begin{document}

\hyphenation{dis-pla-ce-ment re-si-sti-vi-ty ge-ne-ra-ted re-la-ti-vi-stic dif-fe-ren-ce in-sensi-ti-ve ki-ne-ma-ti-cal Sy-ste-ma-tic sy-ste-ma-tic ave-ra-ging con-ver-ters con-ver-ter la-bo-ra-to-ry se-con-da-ry
se-con-da-ries si-mu-la-tion si-mu-la-tions do-mi-na-te do-mi-na-tes ta-bu-la-tions
Ta-bu-la-tions Di-stri-bu-tions
di-stri-bu-tions di-stri-bu-tion Di-stri-bu-tion Di-spla-ce-ment di-spla-ce-ment Di-spla-ce-ments
di-spla-ce-ments Li-near li-near Ca-sca-de Ca-sca-des ca-sca-de ca-sca-des
ta-bu-la-tion Ta-bu-la-tion re-pe-ti-ti-ve E-lec-tron E-lec-trons
e-lec-tron e-lec-trons De-tec-tion Pro-duc-tion pro-duc-tion
Re-so-lu-tions Re-so-lu-tion re-so-lu-tions re-so-lu-tion
Ope-ra-tion mi-ni-mum Ener-gy Ener-gies fer-ro-mag-net
fer-ro-mag-nets meta-sta-ble meta-sta-bi-lity con-fi-gu-ra-tion
con-fi-gu-ra-tions expo-nen-tially mo-bi-li-ty- mo-bi-li-ties
tem-pe-ra-tu-re tem-pe-ra-tu-res con-cen-tra-tion con-cen-tra-tions
elec-tro-nic elec-tro-nics STMelec-tro-nics sec-tion Sec-tion
Chap-ter chap-ter theo-ry ap-pro-xi-mation ra-dia-tion Ra-dia-tion
ca-pa-ci-tan-ce approaches tran-sport dispersion Ca-lo-ri-me-try
ca-lo-ri-me-try En-vi-ron-ment En-vi-ron-ments en-vi-ron-ment
en-vi-ron-ments Fur-ther-mo-re do-mi-nant ioni-zing pa-ra-me-ter pa-ra-me-ters
O-sa-ka ge-ne-ral exam-ple Exam-ple ca-vi-ty Ca-vi-ty He-lio-sphe-re
he-lio-sphe-re dis-tan-ce Inter-pla-ne-ta-ry inter-pla-ne-ta-ry
ge-ne-ra-li-zed sol-ving pho-to-sphe-re sym-me-tric du-ring
he-lio-gra-phic strea-ming me-cha-nism me-cha-nisms expe-ri-mental
Expe-ri-mental im-me-dia-tely ro-ta-ting na-tu-rally
ir-re-gu-la-ri-ties o-ri-gi-nal con-si-de-red e-li-mi-na-ting
ne-gli-gi-ble stu-died dif-fe-ren-tial mo-du-la-tion ex-pe-ri-ments
ex-pe-ri-ment Ex-pe-ri-ment Phy-si-cal phy-si-cal in-ve-sti-ga-ted
Ano-de Ano-des ano-de ano-des re-fe-ren-ce re-fe-ren-ces
ap-pro-xi-ma-ted ap-pro-xi-ma-te in-co-ming bio-lo-gi-cal
atte-nua-tion other others eva-lua-ted nu-cleon nu-cleons reac-tion
pseu-do-ra-pi-di-ty pseu-do-ra-pi-di-ties esti-ma-ted va-lue va-lues
ac-ti-vi-ty ac-ti-vi-ties bet-ween Bet-ween dis-cre-pan-cy
dis-cre-pan-cies cha-rac-te-ri-stic
cha-rac-te-ri-stics sphe-ri-cally anti-sym-metric ener-gy ener-gies
ri-gi-di-ty ri-gi-di-ties leaving pre-do-mi-nantly dif-fe-rent
po-pu-la-ting acce-le-ra-ted respec-ti-ve-ly sur-roun-ding pa-ral-lel
sa-tu-ra-tion vol-tage vol-tages da-ma-ge da-ma-ges be-ha-vior
equi-va-lent si-li-con exhi-bit exhi-bits con-duc-ti-vi-ty
con-duc-ti-vi-ties dy-no-de dy-no-des created Fi-gu-re Fi-gu-res
tran-si-stor tran-si-stors Tran-si-stor Tran-si-stors ioni-za-tion
Ioni-za-tion ini-tia-ted sup-pres-sing in-clu-ding maxi-mum mi-ni-mum
vo-lu-me vo-lu-mes tu-ning ple-xi-glas using de-pen-ding re-si-dual har-de-ning li-quid
know-ledge usage me-di-cal par-ti-cu-lar scat-te-ring ca-me-ra se-cond hea-vier hea-vy trans-axial
con-si-de-ration created Hy-po-the-sis hy-po-the-sis usually inte-ra-ction Inte-ra-ction
inte-ra-ctions Inte-ra-ctions pro-ba-bi-li-ty pro-ba-bi-li-ties
fol-low-ing cor-re-spon-ding e-la-stic readers reader pe-riod pe-riods geo-mag-ne-tic sa-ti-sfac-tory
ma-nu-fac-tu-red ha-zard ha-zards}


\begin{center}
To appear on the Proceedings of the 13th ICATPP Conference on\\
Astroparticle, Particle, Space Physics and Detectors\\ for Physics Applications,\\ Villa  Olmo (Como, Italy), 23--27 October, 2013, \\to be published by World Scientific (Singapore).
\end{center}
\vspace{-1.7cm}

\title{NIEL DOSE DEPENDENCE FOR SOLAR CELLS IRRADIATED WITH ELECTRONS AND PROTONS}

\author{
C. Baur$^{1}$, M. Gervasi$^{2,3}$, P. Nieminen$^{1}$, S. Pensotti$^{2,3}$, \\P.G. Rancoita$^{2,*}$ and M. Tacconi$^{2,3,\ddag}$
}

\address{$^{1}$ESA--ESTEC, Noordwijk, Netherlands\\
$^{2}$Istituto Nazionale di Fisica Nucleare, INFN Milano-Bicocca, Milano (Italy) \\
$^{3}$Department of Physics, University of Milano Bicocca, Milano (Italy)
}
\address{\textbf{*e-mail: piergiorgio.rancoita@mib.infn.it}\\\textbf{$\ddag$ E-mail: mauro.tacconi@mib.infn.it}}

\begin{abstract}
The investigation of solar cells degradation and the prediction of its end-of-life performance is of primary importance in the preparation of a space mission.~In the present work, we investigate the reduction of solar-cells' maximum power resulting from irradiations with electrons and protons.~Both GaAs single junction and GaInP/GaAs/Ge triple junction solar cells were studied.~The results obtained indicate how i) the dominant radiation damaging mechanism is due to atomic displacements, ii) the relative maximum power degradation is almost independent of the type of incoming particle,~i.e., iii) to a first approximation, the fitted \textit{semi-empirical function expressing the decrease of maximum power depends only on the absorbed NIEL dose}, and iv) the actual displacement threshold energy value ($E_d=21\,$eV) accounts for annealing treatments, mostly due to self-annealing induced effects.~Thus, for a given type of solar cell, a unique maximum power degradation curve can be determined as a function of the absorbed NIEL dose.~The latter expression allows one to predict the performance of those solar cells in space radiation environment.
\end{abstract}
\section{Introduction}
Nuclei and electrons populating the heliosphere can induce radiation hazards in space missions.~In fact, these particles can produce atomic displacements in the lattice of semiconductors employed in spacecraft electronics, instrumentation or solar cells.~These displacements lead to permanent damages and, consequently, a failure or degraded performance [1] (see also Chapter 4 in~[2] and references therein).~
\par
The non-ionizing energy loss (NIEL) expresses the amount of energy deposited by an incident particle passing through a material and resulting in displacement processes.~The degradation of semiconductor-device performance is expected to be increased with increasing the amount of absorbed NIEL dose, if the dominant damaging mechanism is caused by atomic displacements.~In this article, we will discuss the radiation damage effects induced in (GaAs) single-junction and (GaInP/GaInAs/Ge) triple-junction solar cells, after irradiations with electrons and protons at different energies.~The results obtained indicated that, within experimental uncertainties, the data are compatible with a unique degradation curve for the solar cell power as a function of the NIEL dose deposited in the device, independently of the type and energy of the incoming particle.
\section{Non-Ionizing Energy-Loss for Electrons and Protons}
\label{NIEL_e_p}
The non-ionizing energy-loss (NIEL) accounts for the amount of energy - lost by a particle passing through a medium -, which was imparted to create atomic displacements.~It can be expressed in units of
MeV per cm, as
\begin{equation}\label{eq2a_NIEL}
    - \left(\frac{dE}{dx}\right)_{\rm NIEL}  \equiv \frac{dE_{\rm de}}{dx},
\end{equation}
where
$\frac{dE_{\rm de}}{dx}$ is the displacement stopping power (e.g.,~see discussion in Sect.~4.2.1 in~[Leroy and Rancoita~(2011)])
and the minus sign for $\left(\frac{dE}{dx}\right)_{\rm NIEL}$ in Eq.~(\ref{eq2a_NIEL}) indicates
that the energy is lost by the incoming particle to create atomic displacements inside the absorber.~For instance, in case of (elastic) Coulomb scattering on nuclei, the displacement stopping power can be calculated as
\begin{equation}\label{eq:NIEL_e_Isotope}
\frac{dE_{\rm de}}{dx} = n_{\rm A}\,\int_{E_d}^{E_{R}^{\max}} \!\!E_{R}\,L(E_{R})\,\frac{d\sigma(E,E_{R})}{dE_{R}}~dE_{R},
\end{equation}
where [see Equation~(1.71) at page~25 of~[2])
\begin{equation}\label{n_A}
n_A = \frac{N \rho_{\rm A}}{A}
\end{equation}
is the number of atoms per
cm$^{3}$ in the absorber, $\rho_{\rm A}$
and $A$ are the density and atomic weight of the
medium, respectively; $N$ is the \index{Avogadro
constant}Avogadro constant, $E$ is the kinetic energy of the incoming particle; $E_{R}$ and ${E_{R}^{\max}}$  are the recoil kinetic energy and the maximum energy transferred to the recoil nucleus, respectively; $E_{d}$ is the so-called \textit{displacement threshold energy}, i.e. the minimum energy necessary to permanently displace an atom from its lattice position; the expression of $L(E_{R})$ is the Lindhard partition function discussed, for instance, in Sects.~4.2.1 and 4.2.1.2 in~[2] (see also references therein and [3]); finally, ${d\sigma(E,E_{R})}/{dE_{R}}$ is the differential cross section for elastic Coulomb scattering for electrons or protons on nuclei.~Furthermore, the non-ionizing energy-loss (NIEL) can be expressed in units of MeV\,cm$^2$/g as
\begin{eqnarray}
\label{eq2a_NIEL_mass} - \left(\frac{dE}{d\chi}\right)_{\rm NIEL}& \equiv & \frac{dE_{\rm de}}{d\chi} \\
\label{eq2a_NIEL_mass1} &=& \frac{N}{A} \,\int_{E_d}^{E_{R}^{\max}} \!\!E_{R}\,L(E_{R})\,\frac{d\sigma(E,E_{R})}{dE_{R}}~dE_{R}
\end{eqnarray}
with $\chi =x \rho_{\rm A}$, $\rho_{\rm A}$ the absorber density in g/cm$^3$ and ${dE_{\rm de}}/{d\chi}$ the displacement mass-stopping power.~
\par
In the current study, the elastic Coulomb cross section of protons on nuclei is the one derived for the treatment of the nucleus--nucleus scattering above 50\,keV/nucleon - up to relativistic energies - and discussed in [4] and Sect.~2.1.4.2 in~[2] (see also [5--7]).~The hadronic contribution to the overall non-ionizing energy deposited can be neglected below 9.5\,MeV - i.e., for the proton energies used in the current investigation -, because, as discussed in [8], this contribution decreases very rapidly with decreasing energy and is lower or equal to $\approx 8.6\%$ at $ 9.5\,$MeV.~Furthermore, the elastic Coulomb interactions of electrons on nuclei is treated - up to relativistic energies (e.g.,~see [9], Sects.~1.3.1--1.3.3 in~[10] and references therein) - using the Mott differential cross section [11] calculated from the practical expression discussed in [12] (see also [13]).
\par
NIEL for compounds can be determined by means of Bragg's rule,~i.e., the overall NIEL [Eq.~(\ref{eq2a_NIEL_mass})] in units of MeV\,cm$^2$/g is obtained as a weighted sum in which each material contributes proportionally to the fraction of its atomic weight.~Thus, for GaAs ones obtains (for instance, see [14] and Equation~(2.20) at page~15 in [15]):
\begin{equation}\label{CompoundGaAs}
\left(\frac{dE_{\rm de}}{d\chi}\right)_{\rm{GaAs}}= \frac{A_{\rm{Ga}}}{A_{\rm{Ga}}+A_{\rm{As}}}\left(\frac{dE_{\rm de}}{d\chi}\right)_{\rm{Ga}}+\frac{A_{\rm{As}}}{A_{\rm{Ga}}+A_{\rm{As}}} \left(\frac{dE_{\rm de}}{d\chi}\right)_{\rm{As}}
\end{equation}
where $\left(\frac{dE_{\rm de}}{d\chi}\right)_{\rm{Ga}}$ $\left[\left(\frac{dE_{\rm de}}{d\chi}\right)_{\rm{As}}\right]$ and $A_{\rm{Ga}}$ [$A_{\rm{As}}$] are the NIEL (in units of MeV\,cm$^2$/g ) and the atomic weight of Gallium [Arsenic], respectively.
\begin{figure}[t]
\begin{center}
 \includegraphics[width=1.0\textwidth]{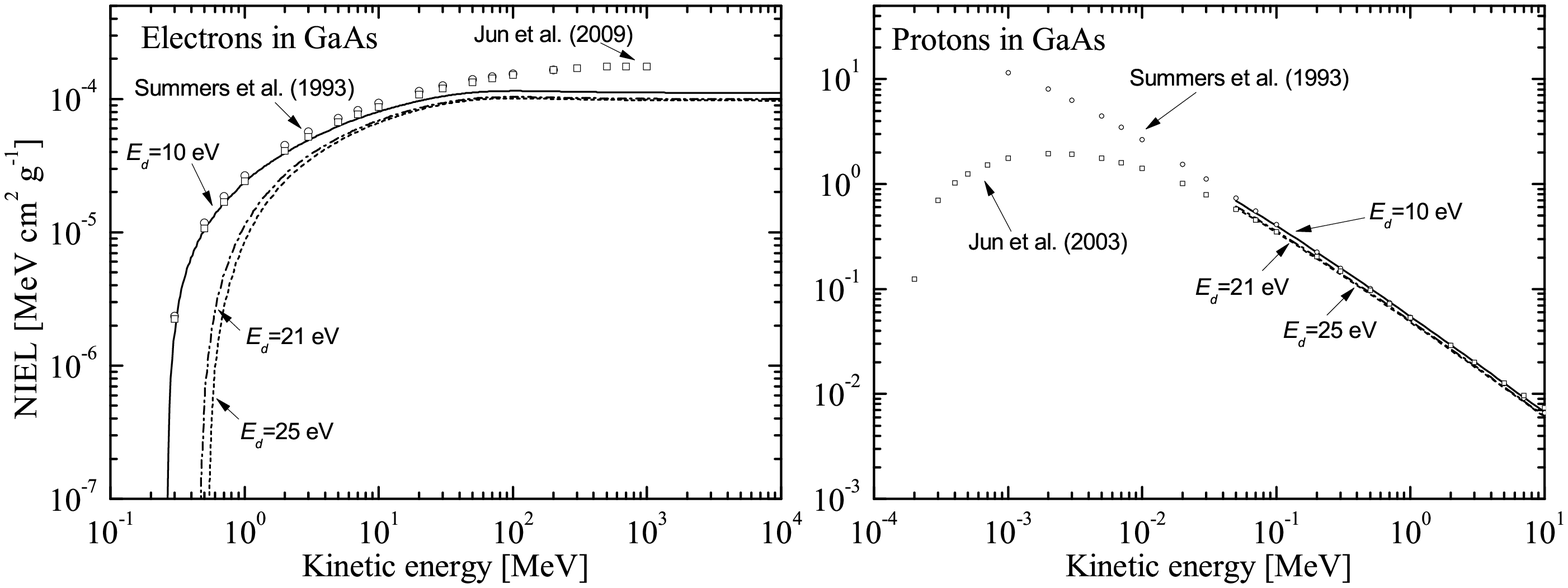}
 \vskip -.3cm
 \caption{
 NIEL in units of MeV\,cm$^2$/g as a function of kinetic energy for electrons (left) from 100\,keV up to 10\,GeV and protons from 0.1\,keV up to 10\,MeV (right) in GaAs.~The calculation for the current work were performed using Eqs.~(\ref{eq2a_NIEL_mass1},~\ref{CompoundGaAs}) with
 $E_d=10,\,21\textrm{ and }25\,$eV; for electrons, data points are from [14, 16] with $E_d=10\,$eV; for protons, data points are from [8, 16] with $E_d=10\,$eV.}
 \label{FiG_niel}
\end{center}
\end{figure}
%
\subsection{Displacement Threshold Energy, Annealing Effects and Absorbed Displacement Dose}
\label{D_T_E_A}
As discussed in Sect.~\ref{NIEL_e_p} - e.g.,~see Eqs.~(\ref{eq:NIEL_e_Isotope},~\ref{eq2a_NIEL_mass1}) -, the displacement stopping power depends on actual values of the displacement threshold energy, $E_d$.~For instance, in Fig.~\ref{FiG_niel} (e.g., see Appendix 1) it is shown the displacement mass-stopping power (i.e.,~the NIEL in units of MeV\,cm$^2$/g) obtained using Eqs.(\ref{eq2a_NIEL_mass1},\,\ref{CompoundGaAs}) with $E_d=10,\,21\textrm{ and }25\,$eV for electrons from 100\,keV up to 10\,GeV and for protons from 0.1\,keV up to 10\,MeV.~In Fig.~\ref{FiG_niel}, for electrons (protons) the data points  are from [14, 16] ([8, 16]) and were obtained with $E_d=10\,$eV.~One can remark that NIEL is almost independent of (or slightly dependent on) $E_d$ above about 50\,keV for protons and (10--20)\,MeV for electrons.~Furthermore, above a few hundred keV's the current calculations are well in agreement with those from [16] and differ only slightly from those in [8].
\par
\begin{figure}[b]
\begin{center}
 \includegraphics[width=1\textwidth]{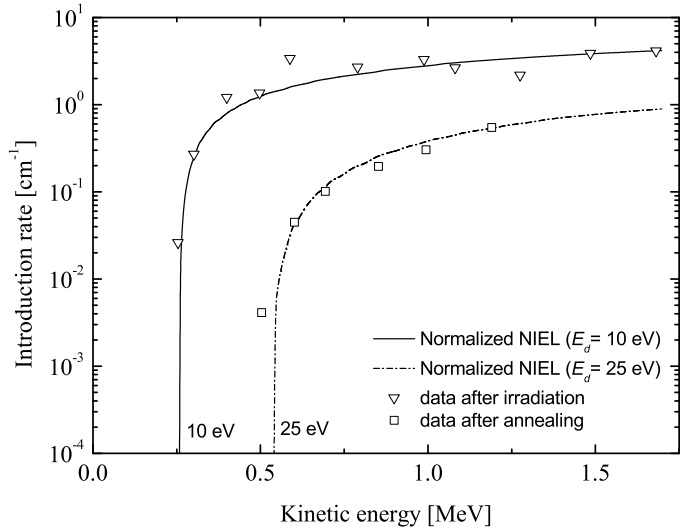}
 \vskip -.4cm
 \caption{Overall introduction rates of defects in GaAs resulting from irradiations with electron as a function of particle energy.~Data after irradiation are from [18], while those ones after annealing come from [17, 18].~The curves are the NIEL values normalized to the highest energy points obtained by means of Eqs.~(\ref{eq:NIEL_e_Isotope},~\ref{eq2a_NIEL_mass1}) with $E_d =10$\,eV (solid line) and 25\,eV (dotted and dashed line).}\label{Figure_Pomms}
\end{center}
\end{figure}

The displacement threshold energy for GaAs was found to be 10\,eV after irradiations with electrons by measuring the overall rate of defects introduced in the bulk of a GaAs diode using DLTS [17, 18] (deep level transient spectrometry).~In addition, those authors provide an experimental evidence that $E_d$ was compatible with 25\,eV when the irradiated samples were annealed at 235\,K (stage I) and 280\,K (stage II).~In Fig.~\ref{Figure_Pomms} the overall introduction rates of defects from [18] and that from [17, 18] obtained after stage I and II annealings are shown; in the figure, the curves correspond to the NIEL calculated by means of Eqs.~(\ref{eq:NIEL_e_Isotope},~\ref{eq2a_NIEL_mass1}) with $E_d =10$\,eV (solid line) and 25\,eV (dotted and dashed line).~The NIEL values (in both cases) are normalized to the highest energy points.
\par
It has to be remarked that the displacement threshold energy was found [17, 18] to be about 10\,eV from the data obtained after irradiations, but annealing even at temperatures lower than $25\,^\circ$C - although close to it - can result in increasing the $E_d$ value, because of the partial recombination regarding point defects (i.e.,~the Frenkel pairs) caused by irradiations.~Thus, the actual displacement threshold energy depends on annealing effects.
\par
The absorbed NIEL dose, $D^{\rm NIEL}$, in units of MeV/g imparted by particles with kinetic energy $E$ can be obtained from (e.g.,~see Equation~4.150 at page~432 in~[2]):
\begin{equation}\label{NIEL_Dose}
D^{\rm NIEL}=\Phi~\frac{dE_{\rm de}}{d\chi}
\end{equation}
where $\Phi$ is the fluence in cm$^{-2}$ of traversing particles (electrons or protons in the current investigations) and $\frac{dE_{\rm de}}{d\chi} $ [Eq.~(\ref{CompoundGaAs})] is the corresponding displacement mass-stopping power due to elastic Coulomb scatterings on nuclei, which - for the incoming particle energies employed for the current study - the only (for electrons) or largely dominant (for protons) physical process resulting in atomic displacements.
%
\section{Single Junction Solar Cells}
\label{Single_J_S_c}
%
\begin{figure}[b]
\begin{center}
 \includegraphics[width=0.9\textwidth]{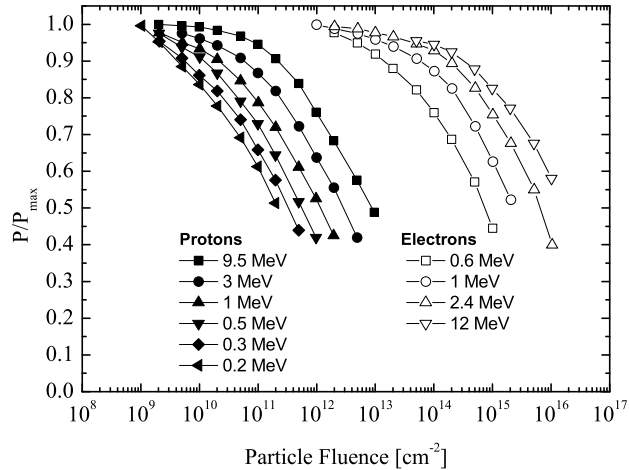}
 \vskip -.3cm
 \caption{P/P$_{\rm max}$ ratio of GaAs solar cells as a function of particle fluence in units of electrons/cm$^{2}$ or proton/cm$^{2}$.}\label{fig_1J_fluences}
\end{center}
\end{figure}
The irradiation data points for single junction solar cells are those from [19], experimentally obtained in [20])(see Fig.~\ref{fig_1J_fluences}).~Those solar cells - produced by Applied Solar Energy Corp. - are GaAs solar cells on an inactive Ge substrate.~Various facilities were used for irradiating those devices with electrons and protons; more information about the solar cell structure, irradiation procedures and facilities can be found in [20].~
\par
In Fig.~\ref{fig_1J_fluences}, the degradation of the maximum power of the solar cells is shown as a function of particle fluence for electrons (with 0.6, 1.0, 2.4 and 12.0\,MeV) and protons (with 0.2, 0.3, 0.5, 1.0, 3.0 and 9.5\,MeV); P/P$_{\rm max}$ is the ratio between the maximum power achieved after irradiation with respect to that one before irradiation,~i.e., it corresponds to the actual maximum power relative ($P_r$) to that before irradiation.~
\par
\begin{figure}[t]
\begin{center}
 \includegraphics[width=1.0\textwidth]{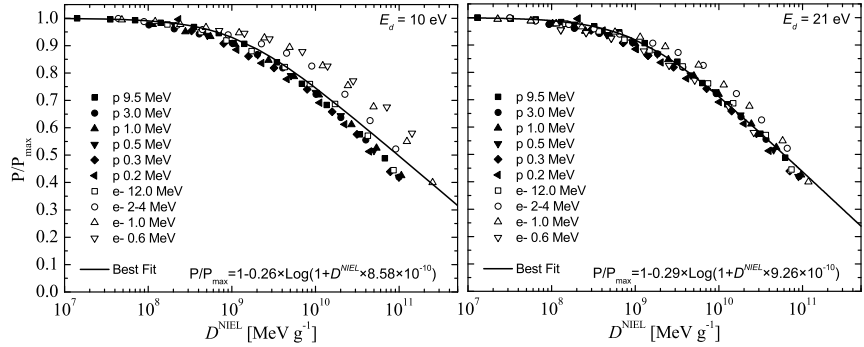}
  \vskip -.2cm
 \caption{P/P$_{\rm max}$ ratio of GaAs single junction solar cells as a function of $D^{\textrm{NIEL}}$ (NIEL doses) obtained from Eqs.~(\ref{eq2a_NIEL_mass1}--\ref{NIEL_Dose}) in MeV/g for $E_d=10\,$eV (left) and $E_d=21$\,eV (right).~The solid line represents the overall best fit of all data points to a semi-empirical function [Eq.~(\ref{semi_emp_fit_eq})].}\label{fig_1J_doses}
\end{center}
\end{figure}

The irradiation fluences can be re-expressed as a function of NIEL doses in MeV/g by means of Eqs.~(\ref{eq2a_NIEL_mass1}--\ref{NIEL_Dose}); thus, the P/P$_{\rm max}$ ratio can be also shown as a function of the NIEL dose (Fig.~\ref{fig_1J_doses}).~In Fig.~\ref{fig_1J_doses}, the solid line represents the best fit for all data to the semi-empirical function [19]:
\begin{equation}\label{semi_emp_fit_eq}
\frac{\rm P}{\rm P_{max}}\equiv P_r=1-C~\textrm{Log}\left(1+\frac{D^{\rm NIEL}}{D_x}\right)
\end{equation}
where $C$ and $D_x$ are parameters to be obtained from the fitting procedure and $D^{\rm NIEL}$ is the NIEL dose, which, in turn, depends on the displacement threshold energy $E_d$.~As discussed in Sect.~\ref{D_T_E_A}, the NIEL dose may depend on annealing effects, for instance those due to self-annealing.~As a consequence, the actual displacement threshold energy can be larger than that  experimentally determined after irradiations [17, 18, 21].~By inspection of Fig.~\ref{fig_1J_doses} (left) in which NIEL doses were calculated for $E_d=10\,$eV, one can notice that a) protons data exhibit a P/P$_{\rm max}$ ratio which depends on the actual NIEL dose and is almost independent on the proton energy -~i.e., as discussed in Sec.~\ref{D_T_E_A}, the NIEL values are almost independent on the actual value of $E_d$ for all proton energies used in the current investigation [e.g.,~see Fig.~\ref{FiG_niel} (right)] -, b) electrons data are off-set with respect to proton data at similar NIEL dose, c) the latter separation decreases with increasing electron energy -~i.e., as discussed in Sect.~\ref{D_T_E_A}, the NIEL values (largely) depend on the actual value of $E_d$ for electron energies lower than 10\,MeV [e.g.,~see Fig.~\ref{FiG_niel} (left)] - and d) the fitted parameters for Eq.~(\ref{semi_emp_fit_eq}) are $C= 0.26$ and $D_x= 1.17\! \times \! 10^{9}\,\textrm{MeV/g}$.~
\par
To account for self-annealing effects - resulting in an increase of the displacement threshold energy -, the NIEL doses were calculated for $E_d$ varying from 10 up to $25\,$eV,~i.e., the values respectively found after irradiation and annealing procedure in [17, 18] and discussed in Sect.~\ref{D_T_E_A}.~For each value of $E_d$, the mean difference (MD) and the square root relative difference (SRRD) of the fitted semi-empirical function [Eq.~(\ref{semi_emp_fit_eq})] with respect to data points were calculated at each electron and proton energy; the quantities MD and SRRD were obtained as:
\begin{eqnarray}
\textrm{MD}=\frac{1}{n_{fl}}\sum^{n_{fl}}_{i=1}\frac{P_{r, i}-P_{fit}}{P_{fit,i}},\\
\textrm{SRRD}=\sqrt{\frac{\sum^{n_{fl}}_{i=1}(P_{r, i}-P_{fit,i})^2}{\sum^{n_{fl}}_{i=1}(P_{fit,i})^2}},
\end{eqnarray}
where
\[
P_{r, i} = \frac{\rm P}{\rm P_{max}},
\]
is the relative maximum power measured at the $i$th fluence (or $i$th corresponding NIEL dose), $P_{fit,i}$ is the relative maximum power of the best-fit curve for the $i$th fluence (or NIEL dose) and $n_{fl}$ is the number of irradiation fluences at a given electron or proton energy.~The optimal fit was obtained using $E_d=21\,$eV, $C= 0.29$, $D_x= 1.08 \! \times \! 10^{9}\,\textrm{MeV/g}$ and is shown in Fig.~\ref{fig_1J_doses} (right).~The corresponding MD$_{op}$ and SRRD$_{op}$ found for $E_d=21\,$eV are listed in Table~\ref{table_1J}.~The SRRD$_{op}$ value averaged over for all energies and particles is 2.5\%; while the $|\textrm{MD}_{op}|$ values do not exceed 3.7\%.
\par
One can state that the best fit obtained with respect to expression~(\ref{semi_emp_fit_eq}) indicates that i) the dominant radiation damaging mechanism is due to atomic displacements, ii) the relative maximum power degradation is almost independent of the type of incoming particle,~i.e., iii) to a first approximation, the fitted \textit{semi-empirical function can be only expressed in terms of the actual NIEL dose absorbed}, and iv) the actual value of $E_d$ is larger that found after irradiation in order to account for the annealing effects (as discussed in Sect.~\ref{D_T_E_A}).
%
\begin{center}
\begin{table}[b]\footnotesize
\tbl{Optimal mean difference (MD$_{op}$) and the square root relative difference (SRRD$_{op}$) to best fitted semi-empirical function [Eq.~(\ref{semi_emp_fit_eq})] for GaAs single junction solar cells as a function of the energy of irradiating electrons and protons [17, 18].~These values were obtained for $E_d=21\,$eV.}
{\begin{tabular}{r|c c|r|c c}
\hline
\multicolumn{ 3}{c|}{Electrons} & \multicolumn{ 3}{c}{Protons} \\ \hline
$E~~~~$ & MD$_{op}$ & SRRD$_{op}$ & $E~~~$ & MD$_{op}$ & SRRD$_{op}$ \\ \hline
0.6 MeV & -2.5\% & 2.7\% & 0.2 MeV & -2.7\% & 3.1\% \\
1.0 MeV & ~2.8\% & 4.2\% & 0.3 MeV & -3.1\% & 3.1\% \\
2.4 MeV & ~3.7\% & 4.4\% & 0.5 MeV & -1.5\% & 1.9\% \\
12.0 MeV & -0.4\% & 2.0\% & 1.0 MeV & -0.9\% & 1.6\% \\
 &  &  & 3.0 MeV & -1.5\% & 1.8\% \\
 &  &  & 9.5 MeV & ~0.0\% & 0.5\% \\ \hline
\end{tabular}}
\label{table_1J}
\end{table}
\end{center}
%
\section{Triple Junction Solar Cells}
\label{Triple_J_S_c}
Triple junction (3J) solar cells under investigation are the 3G28 type manufactured by AZUR SPACE Solar Power GmbH.~These cells are constitued by a GaInP junction at the top, a GaInAs junction at the middle and a Ge junction at the bottom [22--24].~Because of the low Indium concentration, the middle junction characteristics can be approximated to those of a GaAs junction.~Moreover, as discussed for instance in [25], the overall degradation due radiation damage in the 3J solar cell is mostly dominated by that occurring in the GaAs middle junction.~For this reason, the conversion from particle fluences to $D^{\rm NIEL}$ in units of MeV/g was obtained applying the NIEL of GaAs cell for the complete device.
\par
The 3J cells were irradiated with electrons and protons in different facilities and two successive times.~The first irradiation data (``I data-set") were obtained irradiating with i) electrons (1.0 and 3.0\,MeV) at the TU Delft facility (NL) and ii) protons (4.0, 5.0 and 8.2\,MeV) at CEA (France) facility.~The self-annealing took place during (1--2) weeks of storage at room temperature for electrons and protons irradiated samples.~However, for those samples irradiated with 8.2\,MeV protons the storage lasted about (2--3) months at room temperature.
\par
The second set (``II data-set") of cells was irradiated at TU Delft (NL) with electrons (0.5, 1.0 and 3.0\,MeV) and with protons (0.3, 0.75 and 6.5\,MeV) at Isotron (UK), PRL (NL) and CSMSM (France).~In this case, self-annealing of electron irradiated samples took place by keeping both protons and electrons irradiated samples at room temperature for 1 week.~In addition, the samples were annealed for 24 hours at $60\,^\circ$C.
\begin{figure}[t]

\begin{center}
 \includegraphics[width=0.9 \textwidth]{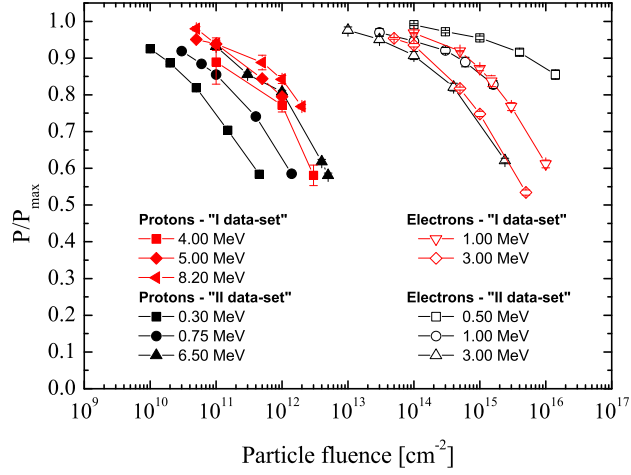}
  \vskip -.2cm
 \caption{P/P$_{\rm max}$ ratio of 3J solar cells as a function of particle fluence in units of electrons/cm$^{2}$ or proton/cm$^{2}$.}\label{Fig_3J_fluences}
\end{center}
\end{figure}
\begin{figure}[b]
\begin{center}
 \includegraphics[width=1.0\textwidth]{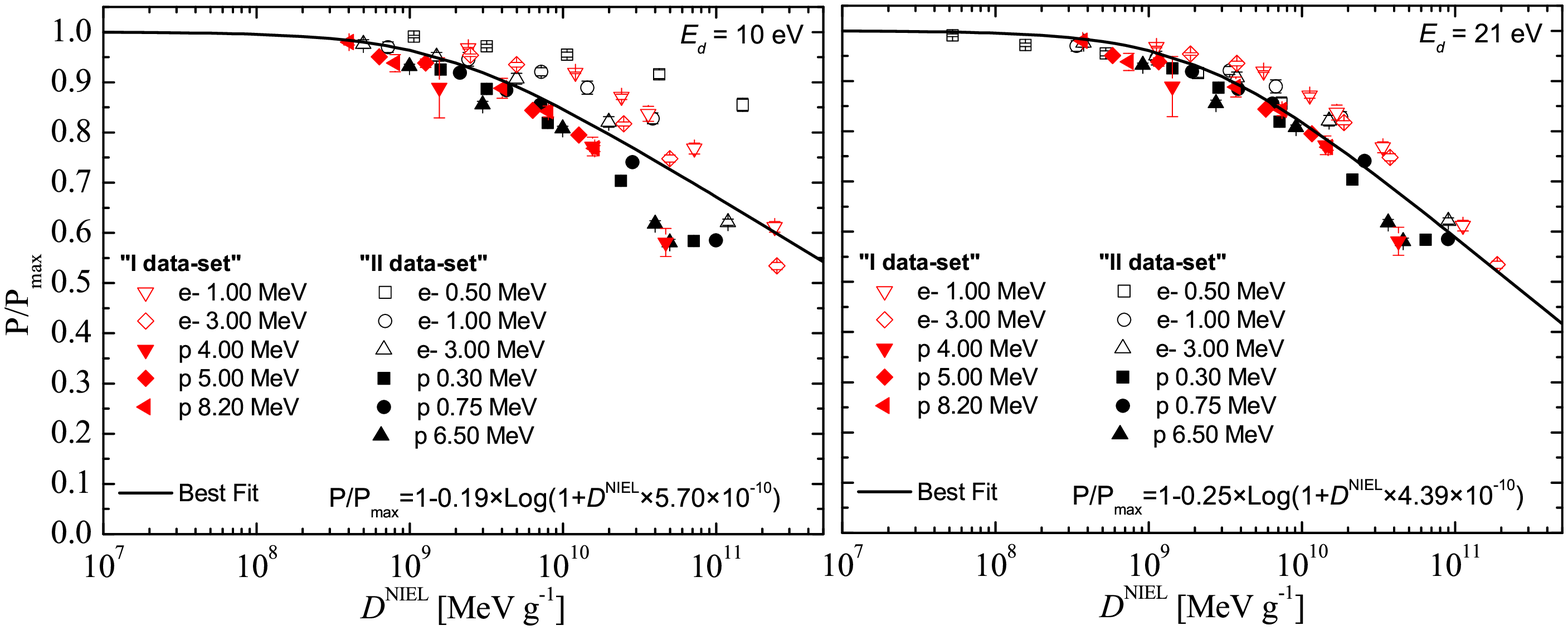}
  \vskip -.2cm
 \caption{P/P$_{\rm max}$ ratio of 3J solar cells as a function of $D^{\textrm{NIEL}}$ (NIEL doses) obtained from Eqs.~(\ref{eq2a_NIEL_mass1}--\ref{NIEL_Dose}) in MeV/g with $E_d=10\,$eV (left), and with $E_d=21\,$eV (right).~The solid line represents the overall best fit of all data points to a semi-empirical function [Eq.~(\ref{semi_emp_fit_eq})].}\label{figure_3J_doses}
\end{center}
\end{figure}
\par
In Fig.~\ref{Fig_3J_fluences}, the P/P$_{\rm max}$ ratios of 3J solar cells are shown as a function of particle fluences for both irradiation data sets.~These ratios are presented as a function of $D^{\textrm{NIEL}}$ (NIEL doses) obtained from Eqs.~(\ref{eq2a_NIEL_mass1}--\ref{NIEL_Dose}) in MeV/g with $E_d=10\,$eV in Fig.~\ref{figure_3J_doses} (left); while in Fig.~\ref{figure_3J_doses} (right), an $E_d=21\,$eV was used for both ``I data-set" and ``II data-set" samples.~In Fig.~\ref{figure_3J_doses}, the solid lines represents the overall best fit of all data points to a semi-empirical function [Eq.~(\ref{semi_emp_fit_eq})]: the fitted parameters for $E_d=21\,$eV ($E_d=10\,$eV) are  $C= 0.25$ (0.19), $D_x= 2.28 \! \times \! 10^{9}\,\textrm{MeV/g}$ ($1.75 \! \times \! 10^{9}\,\textrm{MeV/g}$).~By inspection of Figs.~\ref{figure_3J_doses} (left) and~\ref{figure_3J_doses} (right), one may notice how the change of $E_d$ results in reducing the spread of data points with respect to the semi-empirical function obtained from the fit.~The value of 21\,eV for the displacement threshold energies was found following the same procedure applied to GaAs solar cells and discussed in Sect.~\ref{Single_J_S_c}.~As already remarked in Sects.~\ref{D_T_E_A} and~\ref{Single_J_S_c}, an enhanced value of the displacement threshold energy is expected in order to account for annealing and/or self-annealing effects.~
\par
In Table~\ref{table_3J}, for $E_d=21\,$eV the optimal mean difference (MD$_{op}$) and the square root relative difference (SRRD$_{op}$) to best fitted semi-empirical function [Eq.~(\ref{semi_emp_fit_eq})] are reported; the SRRD$_{op}$ value averaged over for all energies and particles is 4.1\%; while the $|\textrm{MD}_{op}|$ values do not exceed 7.4\%.~The latter values are larger than those found with GaAs single junction solar cells and may account for a) the simplified assumptions made in treating a more complex 3J like a single junction and, possibly, b) the different annealing treatments of the 3J devices with regard to those previously used for single junctions.~
It is worth to remark that SRRD$_{op}$ and $|\textrm{MD}_{op}|$ values do not vary when the Indium content (3\%) is taken into account for the 3J cell.
\par
Finally, as already discussed in Sect.~\ref{Single_J_S_c}, the best fit obtained with respect to expression~(\ref{semi_emp_fit_eq}) confirms the results found with single junction solar cells (Sect.~\ref{Single_J_S_c}) indicating that i) the relative maximum power degradation is almost independent of the type of incoming particle,~i.e., ii) to a first approximation, the fitted \textit{semi-empirical function can be expressed in terms of the absorbed NIEL dose}.
\begin{center}
\begin{table}[t]\footnotesize
\tbl{Optimal mean difference (MD$_{op}$) and the square root relative difference (SRRD$_{op}$) to best fitted semi-empirical function [Eq.~(\ref{semi_emp_fit_eq})] for 3J solar cells as a function of the energy of irradiating electrons and protons for $E_d=21\,$eV.~The ``I data-set" (upper part of the table) samples were irradiated with electrons at 1.0 and 3.0\,MeV and protons at 4.0, 5.0 and 8.2\,MeV; ``II data-set" samples (bottom part of the table) with electrons at 0.5, 1.0 and 3.0\,MeV and protons at 0.3, 0.75 and 6.5\,MeV.}
{\begin{tabular}{r|c c|r|c c}
\hline
\multicolumn{ 3}{c|}{Electrons} & \multicolumn{ 3}{c}{Protons} \\ \hline
$E~~~~$ & MD$_{op}$ & SRRD$_{op}$ & $E~~~~$ & MD$_{op}$ & SRRD$_{op}$ \\ \hline
1.0 MeV & ~6.6\% & 6.8\% & 4.00 MeV & -7.4\% & 8.1\% \\
3.0 MeV&  ~4.9\% & 5.4\% & 5.00 MeV & -1.9\% & 2.0\% \\
 &  &  & 8.20 MeV & -1.3\% & 1.8\% \\
\hline
0.5 MeV & -1.1\% & 1.7\% & 0.30 MeV & -4.5\% & 4.3\% \\
1.0 MeV & ~2.5\% & 4.0\% & 0.75 MeV & -0.8\% & 1.4\% \\
3.0 MeV & ~1.7\% & 2.5\% & 6.50 MeV & -7.2\% & 7.4\% \\
\hline
\end{tabular}}
\label{table_3J}
\end{table}
\end{center}
%
\section{Conclusions}
The degradation of the maximum power in irradiated solar cells with electrons and protons was investigated as a function of particle fluence.~The results obtained from GaAs single junction and GaInP/GaInAs/Ge triple junction solar cells indicate that i) the dominant radiation damaging mechanism is due to atomic displacements, ii) the relative maximum power degradation is almost independent of the type of incoming particle,~i.e., iii) to a first approximation, the fitted \textit{semi-empirical function can be expressed only as a function of the absorbed NIEL dose}, and iv) the actual employed value of $E_d=21\,$eV accounts for the different annealing treatments and, thus, is larger than the $10\,$eV obtained immediately after irradiation using a DLTS technique.~
\par
Furthermore, the measured P/P$_{\rm max}$ ratios agree to the semi-empirical fitted function expressed in terms of the NIEL dose to about some percent, once the appropriate actual displacement threshold energy is used to determine the scale conversion between particle fluence and $D^{\textrm{NIEL}}$.
\par
For the GaAs single junction sola cells, the mean difference (MD) and the square root relative difference (SRRD) to the fitted optimal semi-empirical function were determined.~The SRRD$_{op}$ value is 2.5\%; while the $|\textrm{MD}_{op}|$ values do not exceed 3.7\%, once averaged over all energies and types of particles.~Larger values (4.1\% and 7.4\%, respectively) were found for the 3J junction solar cells and may account for a) the simplified assumptions made in treating a more complex 3J like a single junction to provide the scale conversion between particle fluence and $D^{\textrm{NIEL}}$ and, possibly, b) the different annealing treatments of the 3J devices with regard to those previously used for single junctions.
\section*{Acknowledges}
The work is supported by ASI (Italian Space Agency) under contract ASI-INFN 075/09/0 and ESA (European Space Agency) under ESTEC contract 4000103268.
%
\section*{Appendix 1}
The values of NIEL (Table~\ref{table_electrons_GaAs}) - termed as \textit{screened relativistic NIEL} - in units of MeV\,cm$^2$/g for electrons (see Sect.~\ref{NIEL_e_p}) are obtained using the Mott differential cross section [11] of electrons on nuclei calculated from the practical expression discussed in [12] (see also [13]) and accounting for the effects due to both screened nuclear potentials and form factors discussed in~[9] and Sects.~1.3.1--1.3.3 of~[10] (see also references therein).~
\scriptsize
\begin{center}
\begin{longtable}{r|r|r|r}
\caption{NIEL in units of MeV\,cm$^2$\,g$^{-1}$ for electrons in GaAs as a function of the displacement threshold energy} \label{table_electrons_GaAs} \\
\hline
\multicolumn{ 1}{c|}{\textbf{E (MeV)}} & \multicolumn{1}{c|}{NIEL } & \multicolumn{1}{c|}{NIEL } & \multicolumn{1}{c}{NIEL } \\
\multicolumn{ 1}{c|}{} & \multicolumn{1}{c|}{\textbf{($E_d=10\,$eV)}} & \multicolumn{1}{c|}{\textbf{($E_d=21\,$eV)}} & \multicolumn{1}{c}{\textbf{($E_d=25\,$eV)}} \\ \hline
\endfirsthead

\multicolumn{4}{l}%
{{\tablename\ \thetable{} -- continued from previous page}} \\
\hline
\multicolumn{ 1}{c|}{\textbf{E (MeV)}} & \multicolumn{1}{c|}{NIEL } & \multicolumn{1}{c|}{NIEL } & \multicolumn{1}{c}{NIEL } \\
\multicolumn{ 1}{c|}{} & \multicolumn{1}{c|}{\textbf{($E_d=10\,$eV)}} & \multicolumn{1}{c|}{\textbf{($E_d=21\,$eV)}} & \multicolumn{1}{c}{\textbf{($E_d=25\,$eV)}} \\ \hline
\endhead

\hline
\multicolumn{4}{l}{Continued on next page} \\
\endfoot

\hline
\endlastfoot

2.560E-01 & 3.753E-08 & - & - \\
3.000E-01 & 2.116E-06 & - & - \\
3.500E-01 & 4.499E-06 & - & - \\
4.000E-01 & 7.051E-06 & - & - \\
4.500E-01 & 8.837E-06 & - & - \\
4.620E-01 & 9.190E-06 & 1.246E-08 & - \\
5.000E-01 & 1.061E-05 & 5.231E-07 & - \\
5.260E-01 & 1.169E-05 & 9.265E-07 & 9.935E-09 \\
5.500E-01 & 1.235E-05 & 1.435E-06 & 1.641E-07 \\
6.000E-01 & 1.407E-05 & 2.569E-06 & 8.830E-07 \\
6.500E-01 & 1.537E-05 & 3.609E-06 & 1.718E-06 \\
7.000E-01 & 1.701E-05 & 4.779E-06 & 2.517E-06 \\
7.500E-01 & 1.823E-05 & 5.825E-06 & 3.462E-06 \\
8.000E-01 & 1.939E-05 & 6.915E-06 & 4.527E-06 \\
8.500E-01 & 2.052E-05 & 8.048E-06 & 5.465E-06 \\
9.000E-01 & 2.199E-05 & 9.199E-06 & 6.448E-06 \\
9.500E-01 & 2.304E-05 & 1.007E-05 & 7.478E-06 \\
1.000E+00 & 2.406E-05 & 1.125E-05 & 8.529E-06 \\
1.500E+00 & 3.311E-05 & 1.998E-05 & 1.710E-05 \\
2.000E+00 & 3.979E-05 & 2.698E-05 & 2.397E-05 \\
3.000E+00 & 4.976E-05 & 3.752E-05 & 3.448E-05 \\
4.000E+00 & 5.708E-05 & 4.503E-05 & 4.202E-05 \\
5.000E+00 & 6.318E-05 & 5.092E-05 & 4.794E-05 \\
6.000E+00 & 6.789E-05 & 5.574E-05 & 5.314E-05 \\
7.000E+00 & 7.185E-05 & 5.979E-05 & 5.722E-05 \\
8.000E+00 & 7.526E-05 & 6.328E-05 & 6.072E-05 \\
9.000E+00 & 7.824E-05 & 6.632E-05 & 6.377E-05 \\
1.000E+01 & 8.088E-05 & 6.901E-05 & 6.647E-05 \\
2.000E+01 & 9.713E-05 & 8.550E-05 & 8.303E-05 \\
3.000E+01 & 1.050E-04 & 9.347E-05 & 9.102E-05 \\
4.000E+01 & 1.094E-04 & 9.789E-05 & 9.545E-05 \\
5.000E+01 & 1.119E-04 & 1.004E-04 & 9.800E-05 \\
6.000E+01 & 1.133E-04 & 1.019E-04 & 9.947E-05 \\
7.000E+01 & 1.142E-04 & 1.027E-04 & 1.003E-04 \\
8.000E+01 & 1.146E-04 & 1.032E-04 & 1.007E-04 \\
9.000E+01 & 1.148E-04 & 1.034E-04 & 1.010E-04 \\
1.000E+02 & 1.149E-04 & 1.034E-04 & 1.010E-04 \\
2.000E+02 & 1.139E-04 & 1.025E-04 & 1.001E-04 \\
3.000E+02 & 1.131E-04 & 1.017E-04 & 9.934E-05 \\
4.000E+02 & 1.127E-04 & 1.013E-04 & 9.888E-05 \\
5.000E+02 & 1.124E-04 & 1.010E-04 & 9.858E-05 \\
6.000E+02 & 1.121E-04 & 1.008E-04 & 9.837E-05 \\
7.000E+02 & 1.120E-04 & 1.006E-04 & 9.822E-05 \\
8.000E+02 & 1.119E-04 & 1.005E-04 & 9.810E-05 \\
9.000E+02 & 1.118E-04 & 1.004E-04 & 9.801E-05 \\
1.000E+03 & 1.117E-04 & 1.003E-04 & 9.793E-05 \\
2.000E+03 & 1.114E-04 & 9.999E-05 & 9.759E-05 \\
3.000E+03 & 1.112E-04 & 9.987E-05 & 9.747E-05 \\
4.000E+03 & 1.112E-04 & 9.982E-05 & 9.741E-05 \\
5.000E+03 & 1.111E-04 & 9.978E-05 & 9.738E-05 \\
6.000E+03 & 1.111E-04 & 9.976E-05 & 9.735E-05 \\
7.000E+03 & 1.111E-04 & 9.974E-05 & 9.734E-05 \\
8.000E+03 & 1.111E-04 & 9.973E-05 & 9.732E-05 \\
9.000E+03 & 1.111E-04 & 9.972E-05 & 9.731E-05 \\
1.000E+04 & 1.111E-04 & 9.971E-05 & 9.730E-05 \\
\end{longtable}
\end{center}
\normalsize
\par
The Coulomb contribution to the overall NIEL for protons (Table~\ref{table_protons_GaAs}) above 200\,keV is obtained (see Sect.~\ref{NIEL_e_p}) using the elastic cross section of protons on nuclei derived from treatment of the nucleus--nucleus screened Coulomb scattering discussed in [4] and Sect.~2.1.4.2 of~[2] (see also Refs.~[5--7]).~For proton energies lower than 200\,keV, it was used the 4-terms analytical approximation of the ZBL cross section derived by Messenger et al.~(2004) [see Equations~(1--3, 15) and also references therein] and the Thomas--Fermi screening length (Equation~(2.73) at page~95 of~[2]) as suggested by ICRUM~(1993) for incident protons.~Using such a cross section, the so obtained nuclear stopping powers typically agrees with those found in [15] within a few percent.~For protons, the NIEL contribution resulting from hadronic interactions was estimated by Jun et al. (2003) at $E_d =10\,$eV.~For displacement threshold energies of 21 and 25\,eV it was linearly reduced with respect to one found at 10\,eV by the same amount found for the Coulomb contributions, i.e., by about 7.7 and 9.5\%, respectively.
\scriptsize
\begin{center}
\begin{longtable}{r|r|r|r}
\caption{NIEL in units of MeV\,cm$^2$\,g$^{-1}$ for protons in GaAs as a function of the displacement threshold energy} \label{table_protons_GaAs} \\
\hline
\multicolumn{ 1}{c|}{\textbf{E (MeV)}} & \multicolumn{1}{c|}{NIEL } & \multicolumn{1}{c|}{NIEL } & \multicolumn{1}{c}{NIEL } \\
\multicolumn{ 1}{c|}{} & \multicolumn{1}{c|}{\textbf{($E_d=10\,$eV)}} & \multicolumn{1}{c|}{\textbf{($E_d=21\,$eV)}} & \multicolumn{1}{c}{\textbf{($E_d=25\,$eV)}} \\ \hline
\endfirsthead

\multicolumn{4}{l}%
{{\tablename\ \thetable{} -- continued from previous page}} \\
\hline
\multicolumn{ 1}{c|}{\textbf{E (MeV)}} & \multicolumn{1}{c|}{NIEL } & \multicolumn{1}{c|}{NIEL } & \multicolumn{1}{c}{NIEL } \\
\multicolumn{ 1}{c|}{} & \multicolumn{1}{c|}{\textbf{($E_d=10\,$eV)}} & \multicolumn{1}{c|}{\textbf{($E_d=21\,$eV)}} & \multicolumn{1}{c}{\textbf{($E_d=25\,$eV)}} \\ \hline
\endhead

\hline
\multicolumn{4}{l}{Continued on next page} \\
\endfoot

\hline
\endlastfoot
1.780E-04 & 4.340E-05 & - & - \\
2.000E-04 & 1.211E-01 & - & - \\
2.500E-04 & 4.489E-01 & - & - \\
3.000E-04 & 6.967E-01 & - & - \\
3.500E-04 & 8.930E-01 & - & - \\
3.737E-04 & 9.725E-01 & 3.020E-05 & - \\
4.000E-04 & 1.052E+00 & 5.303E-02 & - \\
4.449E-04 & 1.170E+00 & 2.169E-01 & 1.760E-05 \\
4.500E-04 & 1.183E+00 & 2.346E-01 & 5.756E-03 \\
5.000E-04 & 1.294E+00 & 3.926E-01 & 1.266E-01 \\
5.500E-04 & 1.389E+00 & 5.255E-01 & 2.738E-01 \\
6.000E-04 & 1.468E+00 & 6.426E-01 & 3.998E-01 \\
6.500E-04 & 1.539E+00 & 7.431E-01 & 5.119E-01 \\
7.000E-04 & 1.599E+00 & 8.335E-01 & 6.091E-01 \\
7.500E-04 & 1.652E+00 & 9.126E-01 & 6.960E-01 \\
8.000E-04 & 1.699E+00 & 9.827E-01 & 7.732E-01 \\
8.500E-04 & 1.741E+00 & 1.044E+00 & 8.431E-01 \\
9.000E-04 & 1.776E+00 & 1.101E+00 & 9.042E-01 \\
1.000E-03 & 1.838E+00 & 1.199E+00 & 1.012E+00 \\
1.500E-03 & 1.999E+00 & 1.481E+00 & 1.331E+00 \\
2.000E-03 & 2.041E+00 & 1.597E+00 & 1.470E+00 \\
3.000E-03 & 2.005E+00 & 1.651E+00 & 1.550E+00 \\
4.000E-03 & 1.920E+00 & 1.620E+00 & 1.535E+00 \\
5.000E-03 & 1.827E+00 & 1.564E+00 & 1.491E+00 \\
6.000E-03 & 1.737E+00 & 1.502E+00 & 1.436E+00 \\
7.000E-03 & 1.654E+00 & 1.439E+00 & 1.380E+00 \\
8.000E-03 & 1.577E+00 & 1.380E+00 & 1.326E+00 \\
9.000E-03 & 1.507E+00 & 1.324E+00 & 1.274E+00 \\
1.000E-02 & 1.444E+00 & 1.273E+00 & 1.226E+00 \\
2.000E-02 & 1.029E+00 & 9.210E-01 & 8.920E-01 \\
3.000E-02 & 8.116E-01 & 7.305E-01 & 7.090E-01 \\
4.000E-02 & 6.761E-01 & 6.104E-01 & 5.931E-01 \\
5.000E-02 & 5.827E-01 & 5.270E-01 & 5.125E-01 \\
6.000E-02 & 5.137E-01 & 4.653E-01 & 4.527E-01 \\
7.000E-02 & 4.606E-01 & 4.176E-01 & 4.065E-01 \\
8.000E-02 & 4.183E-01 & 3.795E-01 & 3.695E-01 \\
9.000E-02 & 3.836E-01 & 3.483E-01 & 3.392E-01 \\
1.000E-01 & 3.547E-01 & 3.222E-01 & 3.139E-01 \\
2.000E-01 & 2.159E-01 & 1.949E-01 & 1.899E-01 \\
3.000E-01 & 1.586E-01 & 1.425E-01 & 1.384E-01 \\
4.000E-01 & 1.233E-01 & 1.110E-01 & 1.082E-01 \\
5.000E-01 & 1.015E-01 & 9.151E-02 & 8.900E-02 \\
6.000E-01 & 8.639E-02 & 7.803E-02 & 7.593E-02 \\
7.000E-01 & 7.540E-02 & 6.821E-02 & 6.640E-02 \\
8.000E-01 & 6.698E-02 & 6.047E-02 & 5.908E-02 \\
9.000E-01 & 6.027E-02 & 5.446E-02 & 5.323E-02 \\
1.000E+00 & 5.475E-02 & 4.967E-02 & 4.840E-02 \\
2.000E+00 & 2.927E-02 & 2.670E-02 & 2.614E-02 \\
3.000E+00 & 2.025E-02 & 1.853E-02 & 1.810E-02 \\
4.000E+00 & 1.558E-02 & 1.424E-02 & 1.396E-02 \\
5.000E+00 & 1.279E-02 & 1.174E-02 & 1.148E-02 \\
6.000E+00 & 1.094E-02 & 1.005E-02 & 9.835E-03 \\
7.000E+00 & 9.676E-03 & 8.900E-03 & 8.709E-03 \\
8.000E+00 & 8.758E-03 & 8.043E-03 & 7.891E-03 \\
9.000E+00 & 8.055E-03 & 7.419E-03 & 7.264E-03 \\
1.000E+01 & 7.492E-03 & 6.903E-03 & 6.775E-03 \\
2.000E+01 & 5.241E-03 & 4.843E-03 & 4.746E-03 \\
3.000E+01 & 4.707E-03 & 4.351E-03 & 4.271E-03 \\
4.000E+01 & 4.401E-03 & 4.077E-03 & 3.999E-03 \\
5.000E+01 & 4.160E-03 & 3.857E-03 & 3.783E-03 \\
6.000E+01 & 3.972E-03 & 3.684E-03 & 3.617E-03 \\
7.000E+01 & 3.834E-03 & 3.557E-03 & 3.490E-03 \\
8.000E+01 & 3.735E-03 & 3.466E-03 & 3.401E-03 \\
9.000E+01 & 3.667E-03 & 3.401E-03 & 3.340E-03 \\
1.000E+02 & 3.615E-03 & 3.355E-03 & 3.295E-03 \\
2.000E+02 & 3.324E-03 & 3.087E-03 & 3.034E-03 \\
3.000E+02 & 3.126E-03 & 2.906E-03 & 2.854E-03 \\
4.000E+02 & 3.138E-03 & 2.918E-03 & 2.865E-03 \\
5.000E+02 & 3.281E-03 & 3.051E-03 & 2.995E-03 \\
6.000E+02 & 3.430E-03 & 3.189E-03 & 3.131E-03 \\
7.000E+02 & 3.507E-03 & 3.259E-03 & 3.200E-03 \\
8.000E+02 & 3.503E-03 & 3.255E-03 & 3.197E-03 \\
9.000E+02 & 3.481E-03 & 3.234E-03 & 3.177E-03 \\
1.000E+03 & 3.545E-03 & 3.295E-03 & 3.237E-03 \\
\end{longtable}
\end{center}
\normalsize


\end{document}